\pdfoutput=1

\documentclass[11pt]{article}

\usepackage{acl}

\usepackage{times}
\usepackage{latexsym}
\usepackage{graphicx}
\usepackage[T1]{fontenc}

\usepackage[utf8]{inputenc}

\usepackage{microtype}

\title{A Simple Baseline for Domain Adaptation in End to End ASR Systems Using Synthetic Data}

\author{Raviraj Joshi \\
  Flipkart, Bengaluru \\
  \texttt{raviraj.j@flipkart.com} \\\And
  Anupam Singh \\
  Flipkart, Bengaluru \\
  \texttt{anupam.s@flipkart.com} \\}

\begin{document}
\maketitle
\begin{abstract}
  Automatic Speech Recognition(ASR) has been dominated by deep learning-based end-to-end speech recognition models. These approaches require large amounts of labeled data in the form of audio-text pairs. Moreover, these models are more susceptible to domain shift as compared to traditional models. It is common practice to train generic ASR models and then adapt them to target domains using comparatively smaller data sets. We consider a more extreme case of domain adaptation where text-only corpus is available. In this work, we propose a simple baseline technique for domain adaptation in end-to-end speech recognition models. We convert the text-only corpus to audio data using single speaker Text to Speech (TTS) engine. The parallel data in the target domain is then used to fine-tune the final dense layer of generic ASR models. We show that single speaker synthetic TTS data coupled with final dense layer only fine-tuning provides reasonable improvements in word error rates. We use text data from address and e-commerce search domains to show the effectiveness of our low-cost baseline approach on CTC and attention-based models.   
  
\end{abstract}

\section{Introduction}

End-to-end speech recognition models simplify the speech recognition process by folding multiple components into a single model. These models directly convert the speech utterance into the spoken text \cite{he2019streaming}. The major end-to-end architectures include CTC-based, attention-based, and transducer-based approaches \cite{graves2013speech, graves2012sequence, chan2015listen}. These all neural approaches are competitive in terms of performance however they require a large amount of supervised data to achieve generalization. A model trained on a single application domain doesn't work well on other target domains. Examples of such applications domains include e-commerce, voice search, medical, etc. Since it is not feasible to prepare supervised data for all the application domains, it is common to train models on large out of domain corpus followed by a small amount of in-domain finetuning \cite{bell2020adaptation}. However, this approach still requires the availability of small labeled data. In the most basic form, the unlabelled text data from the target domain can be used to build domain-specific language models (LMs).  The domain LMs are combined with the end-to-end ASR model using shallow fusion \cite{kannan2018analysis,shan2019component,meng2021internal}. This approach has limited benefits since the main ASR model is not tuned to the target domain. Another popular technique is to prepare synthetic data using a Text to Speech (TTS) system and the target domain text data \cite{sim2019personalization}. This requires a sophisticated multi-speaker TTS system followed by the addition of representative noise to make the data usable. The idea is to make synthetic data as close as the real-world data. However, this approach is prone to overfitting as the synthetic data does not exactly resemble real-world noisy conditions. Different fine-tuning approaches have been explored using synthetic data to alleviate the over-fitting problem.

In this work, we are concerned with domain adaptation techniques when a text-only corpus from the target domain is available \cite{gao2021pre}. We present a simple baseline approach using single speaker synthetic TTS data followed by final dense layer only fine-tuning. The synthetic data is created using a single speaker TTS system which is commonly available and also easier to build in-house. The data is not subjected to any noise and is directly used to fine-tune the neural network. Although such single speaker data is easy to build it is not usable for the training of end-to-end networks. We, therefore, propose dense-only fine-tuning for effective fine-tuning. The approach solely relies on final dense layer fine-tuning to avoid over-fitting on single speaker and acoustic conditions. We refer to the dense layer projecting the intermediate embedding onto vocabulary space as the final dense layer. Since the acoustic encoder of the neural network is frozen, the network only learns about the linguistic characteristic of the target domain. Similar approaches have been explored in literature where only the decoder part of the neural network is fine-tuned. However, this approach is not applicable to CTC-based neural networks \cite{graves2006connectionist} which do not follow an encoder-decoder architecture. We present our approach in the context of CTC and Listen-Attend-Spell (LAS) based neural network architectures. For LAS-based network, we also compare dense only and decoder only fine-tuning. We consider the text from address (for delivery of e-commerce products) domain and voice search (of e-commerce products) domain \cite{joshi2021attention} for fine-tuning the model trained on a generic multi-domain dataset. Although encoder only fine-tuning has been widely studied in the literature \cite{mimura2018leveraging}, this is the first work to exploit dense-only fine-tuning which is more relevant to the CTC-based systems. Moreover, we demonstrate a way to build an ASR system for the Address domain which is not explored in the literature. 

\section{Related Work}

Our work is at the intersection of data augmentation using the TTS system and domain adaptation. In this section, we review the recent work in these two areas. The synthetic data generated using the TTS system was used to improve the recognition of out of vocabulary (OOV) words in \cite{zheng2021using}. Both synthetic data containing OOV words and original data were used together to train the best RNN-T model. Encoder freezing and elastic weight consolidation were further shown to provide extra benefits. Similarly, \cite{peyser2019improving} used a TTS system to generate numeric training data and improve the ASR performance on the out of vocabulary numeric sequences. The importance of data augmentation over semi-supervised learning was shown in \cite{laptev2020you}. In this work, the TTS system was trained on the same supervised ASR data set and used to generate synthesized samples on a wider set. The work also highlights the importance of multi-speaker TTS systems and noise addition to build usable systems. Other data augmentation techniques like spec augment \cite{park2019specaugment} were shown to be complementary with TTS based augmentation in  \cite{rossenbach2020generating}. Effective training strategies for using synthetic data were proposed in \cite{fazel2021synthasr}. In order to avoid catastrophic forgetting, multi-stage training was used. The encoder layers were frozen in the initial stage followed by full fine-tuning in later stages. An elastic penalty was also added to the loss function so to avoid large deviation in learned parameters. 

Similar approaches have been proposed for domain adaptation as well with a bais towards fine-tuning based transfer learning approaches. An LSTM-based domain classifier was trained to select an appropriate domain adapted language model in \cite{liu2021domain}. The corresponding domain-specific language model was used for second pass re-scoring. The transfer learning approaches for domain adaptation and cross-language adaptation were evaluated in \cite{huang2020cross}. They compare the fine-tuning of the pre-trained QuartzNet model with the corresponding model trained from scratch. They concluded that large pre-trained models performed better than small pre-trained models and the models trained from scratch. Another form of transfer learning involves partial fine-tuning of the model instead of the entire model. The decoder only fine-tuning for domain adaptation in Listen-Attend-Spell (LAS) based model was evaluated in \cite{ueno2018encoder}. The model is first trained on the source domain followed by decoder only fine-tuning on the target domain. The partial fine-tuning is shown to work better than the full fine-tuning and from the models trained from scratch. An adaptation technique specific to RNN-T networks using text-only data was proposed in \cite{pylkkonen2021fast}. The prediction network of RNN-T is viewed as a neural language model and is adapted using text-only corpus while keeping the encoder and joint network fixed. Another approach for adapting RNN-T network using text-only data was proposed in \cite{li2020developing}. The fine-tuning of prediction and the joint network was performed using synthetic TTS domain-specific data. Partial fine-tuning was shown to work better than full fine-tuning approaches.
These works mainly used RNNT-based systems and employ a multi-context multi-speaker TTS system. In this work, we use a single speaker TTS system with a focus on CTC and attention-based models. Moreover, we focus on dense only fine-tuning instead of decoder fine-tuning studied in these works. 

\section{Methodology}
\begin{figure}
  \centering
  \includegraphics[width=1\columnwidth]{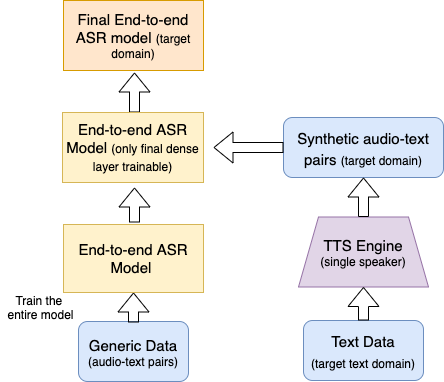}
  \caption{Domain Adaptation Process}
  \label{fig:domain_adapt}
\end{figure}

The flow of our process is depicted in Figure \ref{fig:domain_adapt}. We follow a simple pre-training and fine-tuning approach. The model is first trained on general out-of-domain data. The target domain text data is converted into audio using a single speaker TTS engine. The synthetic samples are then used to fine-tune the final dense layers of ASR models. We consider two model types i.e CTC based models and Attention-based models. The model architecture and TTS system description are provided in the following sub-sections. 
\subsection{Model Architecture}
We consider CTC and attention-based ASR models which follow the same pre-processing steps \cite{joshi2021attention}. The audio is segmented into 20ms chunks with an overlap of 10ms. Log-mel features are computed and provided as input to the model. Standard spec-augment is used for time and frequency masking of the spectrograms \cite{park2019specaugment}. An 80-dimensional log mel feature is computed per time step. Three consecutive features are stacked to give a final feature of size 240. The output vocabulary size consists of sub-word units of size 5000. The sentence piece library is used to train the subword model using the generic out-of-domain data \cite{kudo2018subword}. 

The CTC-based model consists of a series of stacked LSTM layers followed by a final dense layer projecting the hidden vectors onto the vocabulary space. The LSTM consists of 700 units at all levels. A total of 12 LSTM layers are present with a final dense layer of size 700 x 5001. The final vocabulary element is reserved for the blank token. The CTC loss function is used to train the model.

The attention-based model follows a transformer LAS architecture. It consists of 10 encoder layers and 2 decoder layers. All the layers are standard transformer blocks. The internal model dimension is 512 units and the feed-forward dimension is 2048 units. Each block has 4 attention heads with 1024 units each. The final dense layer on the decoder side has a size of 512 x 5000. The same generic vocabulary is used for all the experiments. This sequence to sequence model is trained using the cross-entropy loss function.

A single speaker TTS system is used to generate the synthetic data. The system is based on Tacotron2 architecture \cite{shen2018natural} and a Clarinet \cite{ping2018clarinet} based vocoder. The Tacotron2 sub-system converts a sequence of phonemes to a mel-spectrogram. The generated mel-spectrogram is converted into a time-domain using a Clarinet-style vocoder. In-house single speaker studio recordings are used to train this model. Both Hindi and English queries were recorded using the voice of the same artist and the text was represented in Devnagari script. The TTS system could therefore be used to convert both English and Hindi text to audio.

\begin{table*}
  \centering
  \begin{tabular}{ccccl}
    \hline
    \textbf{Model} & \textbf{Test WER} & \vtop{\hbox{\strut \textbf{Test WER} + }\hbox{\strut \textbf{LM Rescoring}}} & \textbf{N-Best WER} \\
    \hline
    LAS-Gen & 25.31 & 22.18 & 13.71\\ \hline
    LAS-Dense& 16.25 & 15.55 & 7.6\\ \hline
    LAS-Decoder & 13.65 & \textbf{13.36} & 5.82\\ \hline
    CTC-Gen & 31.84 & 25.58 & 13.83 \\ \hline
    CTC-Dense & 20.32 & 17.66 & 8.24\\ \hline
\end{tabular}
\caption{Word Error Rate(WER) for different model variations using Voice Search Domain. The N-Best WER indicates the best WER in the top N=10 beams.}
\label{tab:wer_vs_domain}
\end{table*}

\begin{table*}
  \centering
  \begin{tabular}{ccccl}
    \hline
    \textbf{Model} & \textbf{Test WER} & \vtop{\hbox{\strut \textbf{Test WER} + }\hbox{\strut \textbf{LM Rescoring}}} & \textbf{N-Best WER} \\
    \hline
    LAS-Gen & 39.42 & 31.62 & 25.35\\ \hline
    LAS-Dense& 22.57 & 16.38 & 11.01\\ \hline
    LAS-Decoder & 18.96 & \textbf{12.54} & 8.17\\ \hline
    CTC-Gen & 31.08 & 22.81 & 19.74 \\ \hline
    CTC-Dense & 22.43 & 15.42 & 12.15\\ \hline
\end{tabular}
\caption{Word Error Rate(WER) for different model variations using Address Domain. The N-Best WER indicates the best WER in the top N=10 beams.}
\label{tab:wer_address_domain}
\end{table*}

\subsection{Dataset Details}
The train data consists of a multi-domain generic audio corpus and two domain-specific synthetic data sets. The generic data consists of crowdsourced read speech corpus. It consists of around 4 million samples amounting to 6500 hours of data. The domain-specific data were synthetically created using a single speaker TTS engine. The two domains under consideration are the voice search domain and address domain. The search domain corresponds to the Flipkart e-commerce product search domain. The address domain corresponds to the pan India delivery address.  Both the domains contain around 3 million samples which are approximately 4000 hours of data for VS domain and 5000 for the address domain. The address domain queries are longer as compared to voice search queries. All the datasets consist of queries in both English and Hindi. All the text is represented in the Devanagari script. The test data was real-world domain-specific data recorded on the Flipkart application. The test data is multi-speaker data recorded in a noisy environment and it is very different than the single speaker TTS data recorded in noise-free studio settings. The test audio data was manually transcribed by the operations team. The voice search test data consisted of 25000 examples and address test data had 7000 examples. Except for linguistic overlap, the synthetic train and real test datasets represent completely different environments, and hence improvements reported in this work are not dependent on the quality of the TTS system as long as it is a single speaker.

\section{Results}
In this work, we evaluate dense only fine-tuning baseline for CTC and attention-based models. The domain adaptation approach is presented on two datasets from voice search and address domain. The word error rates(WER) is used to compare the different approaches. The WER is word-level Levenshtein distance between ground truth text and output text. The results for voice search domain and address domain are shown in Table \ref{tab:wer_vs_domain} and Table \ref{tab:wer_address_domain} respectively. The models are first trained on the generic multi-domain dataset and represented as CTC-Gen and LAS-Gen. These pre-trained models are then fine-tuned single speaker synthetic dataset. We show that dense only fine-tuning provides considerable improvement in accuracy while at the same time avoiding over-fitting on single speaker data. The dense-finetuned models are referred to as CTC-Dense and LAS-Dense. We also evaluate decoder-only fine-tuning for LAS models termed as LAS-Decoder. We report WER with and without external language model rescoring. A kenLM based language model is trained using text transcripts for both the domains individually. The N-Best WER is computed by picking the best beam from the top N=10 beam elements. 

The results show that LAS-Dense provides around 30\% relative improvement in WER over LAS-Gen for VS domain and around 50\% relative improvement for the address domain. The LAS-Decoder further improves the results by 14\% for VS domain and 23\% for the address domain. Similarly, CTC-Dense provides an improvement of 30\% and 32\% for VS and address domain respectively over CTC-Gen. Note that the WER of LAS-Gen evaluated on address domain is considerably high as compared to VS domain. Moreover, this simple fine-tuning and LM-rescoring provides high improvements in WER. This shows that the text distribution of address data is very different from the initial multi-domain data. Also, the variety of named entities is very high in address data as compared to VS data. Overall we show that dense only fine-tuning can provide us a reasonable baseline for domain adaptation. For encoder-decoder architectures, decoder fine-tuning serves as a better option. This is expected as the encoder part can also be seen as the acoustic network is frozen and the decoder network which can be seen as a contextual language model is fine-tuned. For CTC-based networks, we observe that extending fine-tuning to even a single lower LSTM layer results in over-fitting and degradation in performance. Therefore for CTC networks dense only fine-tuning is the optimal approach to avoid over-fitting.


\section{Conclusion}
In conclusion, we demonstrate a simple baseline approach for domain adaptation using a text-only corpus from the target domain. We show that the final dense layer only fine-tuning using single speaker TTS data provides considerable improvements over the generic model. The results are shown on two different domains of voice search and address domain. For both CTC and attention-based models we show that dense-only fine-tuning is a reasonable approach for domain adaptation. Although the technique is more relevant to CTC-based models it can also be used with encoder-decoder type models. For encoder-decoder models, the decoder only fine-tuning performs better.

\bibliography{main}
\bibliographystyle{acl_natbib}

\appendix



\end{document}